\documentclass[final,5p,times,twocolumn]{elsarticle}
\usepackage{natbib}
\usepackage{amssymb}
\usepackage{amsmath}
\usepackage{amsfonts}
\usepackage{color}
\usepackage{epstopdf}
\usepackage{hyphenat}
\usepackage{xcolor}
\usepackage{lipsum}
\usepackage{hhline}

\usepackage{array}

\makeatletter
\newcommand{\thickhline}{%
	\noalign {\ifnum 0=`}\fi \hrule height 1.5pt
	\futurelet \reserved@a \@xhline
}
\newcommand{\highlight}[1]{\color{black}#1}

\newcolumntype{"}{@{\hskip\tabcolsep\vrule width 1.5pt\hskip\tabcolsep}}
\makeatother
\begin{document}
	
	\begin{frontmatter}
		
		%%\title{\tnoteref{title}}
		
		\author{Eero Satuvuori\corref{cor1}\fnref{label1,label2,label3}}
		\ead{eero.satuvuori@gmail.com}   
		\author{Thomas Kreuz\corref{cor1}\fnref{label1}}
		\ead{thomas.kreuz@cnr.it}

		\fntext[label1]{Institute for Complex Systems, CNR, Sesto Fiorentino, Italy}
		\fntext[label2]{Department of Physics and Astronomy, University of Florence, Sesto Fiorentino, Italy}
		\fntext[label3]{MOVE Research Institute, Department of Human Movement Sciences, Vrije Universiteit Amsterdam, The Netherlands}
		
		%spike train (synchrony), names, burst, generalizing, adapting... to data with bursts 

\title{Which spike train distance is most suitable for distinguishing rate and temporal coding?}


\begin{abstract}

\noindent \textit{Background:} It is commonly assumed in neuronal coding that repeated presentations of a stimulus to a coding neuron elicit similar responses. One common way to assess similarity are spike train distances. These can be divided into spike-resolved, such as the Victor-Purpura and the van Rossum distance, and time-resolved, e.g. the ISI-, the SPIKE- and the RI-SPIKE-distance. 

\noindent \textit{New Method:} We use independent steady-rate Poisson processes as surrogates for spike trains with fixed rate and no timing information to address two basic questions: How does the sensitivity of the different spike train distances to temporal coding depend on the rates of the two processes and how do the distances deal with very low rates?

\noindent \textit{Results:} Spike-resolved distances always contain rate information even for parameters indicating time coding. This is an issue for reasonably high rates but beneficial for very low rates. In contrast, the operational range for detecting time coding of time-resolved distances is superior at normal rates, but these measures produce artefacts at very low rates. The RI-SPIKE-distance is the only measure that is sensitive to timing information only.

\noindent \textit{Comparison with Existing Methods:} While our results on rate-dependent expectation values for the spike-resolved distances agree with \citet{Chicharro11}, we here go one step further and specifically investigate applicability for very low rates.

\noindent \textit{Conclusions:} The most appropriate measure depends on the rates of the data being analysed. Accordingly, we summarize our results in one table that allows an easy selection of the preferred measure for any kind of data.

\end{abstract}


\end{frontmatter}

\section{Introduction}

Neurons respond to stimulation with discrete events called spikes and a consecutive sequence of such spikes over time form a spike train \citep{Rieke96}. The effect of a spike is to release neuronal transmitters at synapses, and with enough input from the other neurons a downstream neuron fires a spike \citep{Bear07}.

Neurons can code information in the rate and/or in the timing of the spikes \citep{QuianQuiroga13}. 
Neuronal information whether coded in spike rate or in spike timing is linked to the length of the encoding window. The rate at which a relevant property of the neural code can change, and the maximum rate at which changes in the stimulus can be represented, is the rate at which the neural code can adjust to the new representation. A definition for rate and time coding schemes has been provided by \citet{Theunissen95} as follows: 

\begin{itemize}
	
	\item \textit{"It is generally accepted that a rate encoding
		scheme is one in which the relevant information
		encoded about the stimulus is correlated only with the
		number of elicited spikes within the encoding window
		and is not correlated with any aspect of the temporal
		pattern of the spikes within the encoding window."}
	
	\item \textit{"In a temporal encoding scheme, the relevant information
		is correlated with the timing of the spikes within the
		encoding window, over and above any information that
		might be correlated with the number of spikes within
		the window."} 
	
\end{itemize}

\highlight{It is important to note that this definition of temporal coding does not require rate correlation actually to be present. Also it is not limited to single spike correlations but applies to any correlations in spike patterns that would not be expected due to rate alone.}

There is plenty of evidence suggesting that information is coded in rate averaged over short periods of time. An example is rate coding in directional tuning within the motor cortex \citep{georgopoulos95}. However, another view is that for very fast reaction sensory and motor systems this kind of temporal averaging is too slow \citep{vanrullen05}. 
When the organism needs to make decisions based on sensory stimuli in time scales of hundreds of milliseconds it is not possible to average over time at every stage of the pathway from stimulus detection to reaction. Thus it has been proposed that for each neuron first the spike latency of a response could code for intensities of a stimulus forming tuning curves similar to those of rate \citep{vanrullen05}. The problem is that there is no specific trigger instant from when to calculate time delays.  
However, spike time coding has been shown to be very generally used in relaying among others gustatory, somatosensory, olfactory, auditory, and visual information \citep{DiLorenzo03, Harvey13, MacLeod98, Fukushima15, Reich01} at least as long as relevant timing signals are available to the experimenter \citep{vanrullen05}. Using a simulation of ganglion cells it could be demonstrated that information transfer from multiple cells firing only one single spike can exceed that of pure rate coding \citep{vanRullen01}.

Whichever coding type is used, the main assumption behind all neuronal coding research is that repeated presentations of the same stimulus result in a similar spike train responses, whereas presentations of different stimuli typically yield very dissimilar responses.
One very common approach to measure such similarity are spike train distances (see e.g. \citeauthor{houghton10}, \citeyear{houghton10}; \citeauthor{victor15}, \citeyear{victor15}), 
which are designed to assess similarity based on rate and timing within spike trains.
In this paper we deal with four commonly used as well as the more recent spike train distances. The spike-resolved Victor-Purpura distance \citep{Victor96,Victor97}, and the van Rossum distance \citep{VanRossum01} are defined by using spikes as the main elements of the measures while the time-resolved ISI-distance \citep{Kreuz07c,Kreuz09}, the SPIKE-distance \citep{Kreuz11,Kreuz13}, and the RI-SPIKE-distance \citep{Satuvuori17} are based on time.

The Victor-Purpura distance \citep{Victor96,Victor97} and the van Rossum distance \citep{VanRossum01} utilize time scale parameters $q$ and $\tau$ respectively. 
For the extreme time scale parameter values $q = 0$ and $\tau$ approaching infinity, respectively, the distances evaluate spike count as an indicator of rate. While this works in some cases, it ignores any temporal correlations in spike timing within the counting window. In order to be more and more sensitive not only to rate but also to timing information, the time-scale parameter has to be moved towards the opposite side of the parameter range.
For this reason, these parameters are often taken as deciding between rate or time coding \citep{Victor05}.
Spike-resolved spike train distances have been applied to identify rate and time coding in recordings e.g. from the auditory system (in songbirds, \citeauthor{wang07}, \citeyear{wang07}; \citeauthor{tang14}, \citeyear{tang14}) or the visual system (salamander retinas, \citeauthor{Chichilnisky05}, \citeyear{Chichilnisky05}, or visual cortex of cats, \citeauthor{Samonds04}, \citeyear{Samonds04}).
On the other hand, the time-resolved ISI-distance \citep{Kreuz07c,Kreuz09}, SPIKE-distance \citep{Kreuz11,Kreuz13}, and RI-SPIKE-distance \citep{Satuvuori17}, are time scale free and thus do not require a parameter, but they assess timing information in the data in relation to the local time scale.
 
The existence of rate and time coding in neural representations within the brain (or even their coexistence in different brain regions) is a very important field of investigation. Since spike train distances are used to quantify the (dis)similarity of spike trains are thus one of the most used tools to tackle this issue, it is equally important to understand their definitions of similarity. Therefore, in the first part of this study we investigate the specific sensitivities of the measures to rate and time coding. Any distance responding to rate difference between the spike trains is sensitive to rate coding. If a distance is sensitive to spike timing, it causes deviations from the distance obtained for pure rate coding (rate difference) if such can be defined for the measure. This definition also implies that if a timing responsive distance is not sensitive to rate difference, it assesses purely timing information. These distinctions allow us to quantify the operational ranges of the distances for detecting time coding using pairs of independent steady rate Poisson spike trains (with reasonably high rates) as surrogates for random spike trains with no timing information. 

In the second part of this study we look at limitations of the different spike train distances regarding their ability to deal with very low spike rates in the data. Neuronal spike trains have a very strong restriction regarding the information they can convey over the spike generating process that is called the floor effect. This is because they are discrete samples from a distribution and the smallest units measurable is a single spike. As an example one might have a neuron firing at 1Hz for a 1s recording and the spike count distribution consists mostly of values 0, 1, or 2 spikes. Equally, if a neuron fires at 2Hz, it still exhibits a considerable amounts of spike trains with 0, 1 or 2 spikes, even if it is more likely to also produce spike trains with 4 or 5 spikes. Since differences between 10Hz and 20Hz are more visible from fewer recordings, this floor effect only affects the low end of the rate scale. Here, due to the insufficient sampling it is not possible to carry out a meaningful statistical analysis of the spike train distances based on the number of spikes. Instead, in this part of the study we use sampling over multiple realizations of steady rate Poisson processes to estimate the minimal rate that is needed in order to still obtain reliable estimates of timings in the data.

The remainder of the paper is organized as follows. In the Methods (Sec. \ref{s:Methods}) we introduce the spike train distances used in this study and introduce the statistical method used. The Results (Sec. \ref{s:Results}) are divided into three parts. In the first part we analyze both the spike-resolved and the time-resolved distances for the normal case where the total rate of the processes is reasonably high and thus far from exhibiting the floor effect. In the second part we examine the functionality of both types of distances for rates so very low that the floor effect takes place. A series of simple examples specifically constructed to illustrate the most important implications of the results obtained is presented in the third part. Finally, in Sec. \ref{s:D&C} we discuss the results and present our conclusions. 

\section{\label{s:Methods}Methods}

There are many different ways to quantify similarity in spike trains. In this Section we qualitatively describe four established and one recently proposed spike train distances. These spike train distances can be divided into two main categories. The first category (Sec. \ref{ss:Methods-Spike}) contains measures where spikes are the main element for constructing the distance. For the second category (Sec. \ref{ss:Methods-Time}) the main unit of the analysis is time since the values are assigned over time rather than over spikes. The mathematical definitions for the spike-resolved and time-resolved spike train distances can be found in \ref{App:SpikeResolved} and \ref{App:TimeResolved}, respectively. Finally, in Sec. \ref{ss:Methods-Analysis} we summarize the analysis and the statistical methods used in this study.

\subsection{\label{ss:Methods-Spike}Spike-resolved spike train distances}

The spike-resolved distances are based on the idea that for each spike there should be a matching spike in the other spike train. Thus even if not defined exactly in this manner, they will try finding pairs for spikes and consider unpaired spikes as rate difference. Thus, they consider rate first and timing second, since any excess spikes are always considered in the distance and only after that the timing differences are added.

Here we describe the two spike-resolved distances proposed by Victor and Purpura and by van Rossum which so far have been most commonly used in the context of neuronal coding. Other spike-resolved measures such as SPIKE-synchronization \citep{Kreuz15} are not considered in the scope of this study.
  
\subsubsection{\label{sss:VP}Victor-Purpura distance}

The Victor-Purpura distance \citep{Victor96,Victor97} is based on finding the minimal cost for transforming one spike train into the other using the three elementary operations of deleting, inserting and shifting spikes. 
While both deletion and insertion carry a cost of 1, the parameter $q$ is used to evaluate the cost of shifting a spike. The theoretical range of values the Victor-Purpura distance can obtain for a pair of spike trains will always fall in range of $[n_2-n_1, n_2+n_1]$, where $n_1$ and $n_2$ are the spike counts of the two spike trains. With a $q$-value of zero, shifting spikes to coincide costs nothing and in this case the total distance between spike trains is just the difference in spike count ($n_2-n_1$), since the extra spikes in the spike train with the higher rate need just to be deleted to convert one to another. On the other hand, very high $q$ values require more timing accuracy between spikes, since the distance is the sum of all spikes in the spike trains that do not exactly match a spike in the other spike train (up to $n_2+n_1$). Thus the parameter $q$ is often taken as an indicator of the relative importance of rate and time coding \citep{Victor05}.

  \begin{table*}[t]
  	\begin{center}
  		\begin{tabular}{ | c |c| c|c " c| c |c |}
  			\hline
  			
  			Distance & \multicolumn{3}{| c "}{ Reasonably high rates } & \multicolumn{3}{ c|}{ Very low rates }\\ \cline{2-7}
  			
  			& Timing & Rate & Synchrony & Timing & Rate & Synchrony \\  \hline
  			
  			Victor-Purpura&\textcolor{red}{x } (\textcolor{green}{\checkmark}*)&\textcolor{green}{\checkmark\checkmark}&\textcolor{red}{x } (\textcolor{green}{\checkmark}*)&\textcolor{green}{\checkmark}&\textcolor{green}{\checkmark}&\textcolor{green}{\checkmark}\\ \hline
  			
  			van Rossum&\textcolor{red}{x } (\textcolor{green}{\checkmark}*)&\textcolor{green}{\checkmark}&\textcolor{red}{x } (\textcolor{green}{\checkmark}*)&\textcolor{green}{\checkmark}&\textcolor{green}{\checkmark}&\textcolor{green}{\checkmark}\\ \thickhline

  			ISI&\textcolor{green}{\checkmark}&\textcolor{green}{\checkmark}&\textcolor{green}{\checkmark\checkmark}&\textcolor{red}{x}&\textcolor{red}{x}&\textcolor{red}{x}\\ \hline
  			
  			SPIKE&\textcolor{green}{\checkmark}&\textcolor{green}{\checkmark}&\textcolor{green}{\checkmark\checkmark}&\textcolor{red}{x}&\textcolor{red}{x}&\textcolor{red}{x}\\\hline
  			
  			RI-SPIKE&\textcolor{green}{\checkmark\checkmark}&\textcolor{red}{x}&\textcolor{green}{\checkmark}&\textcolor{red}{x}&\textcolor{red}{x}&\textcolor{red}{x}\\\hline

  			\multicolumn{7}{| l |}{ \textcolor{green}{\checkmark}\textcolor{green}{\checkmark}: very well suited}\\
  			\multicolumn{7}{|l |}{\textcolor{green}{\checkmark}: can perform }\\
  			\multicolumn{7}{| l |}{  \textcolor{red}{x}: avoid using }\\
  			\multicolumn{7}{| l |}{  *: timing information available only for near equal rates }\\
  			\hline

  		\end{tabular}
  		
  	\end{center}
  	
  	\caption[Table]{An overview of all spike train distances. Reasonably high rates denotes the normal case of rates that are high enough to avoid any floor effect (more than 4 spikes overall), whereas very low rates refers to rates so low that the floor effect takes place (total number of spikes equal to or lower than 4). Timing means pure spike timing information, rate pure spike count difference, and synchrony taking into account time local similarities in both rate and spike timing.
  		\label{Tbl:All_distances}}
  \end{table*}
   
\subsubsection{van Rossum distance}

The van Rossum distance \citep{VanRossum01} ties the spikes to a more biological context by using a kernel, which can be considered as the effect a spike will have on the postsynaptic neuron (see also \citeauthor{Houghton09}, \citeyear{Houghton09}). An exponentially decaying kernel with time constant $\tau$ is applied to each spike and differences in the effect patterns are considered when calculating the distance. The larger is $\tau$, the longer lasts the effect of a spike. 

Although the resulting profiles are time-dependent there is no time-normalization. Rather, the values are obtained in a spike-based manner, since each spike is convolved with a kernel function. For this reason the distance obtained from the time-resolved profile is actually spike-resolved. Often the Victor-Purpura distance and the van Rossum distance are considered interchangeable with parameter conversion $\tau = 1/q$ (compare \citeauthor{Schrauwen07}, \citeyear{Schrauwen07}).

\subsection{\label{ss:Methods-Time}Time-resolved spike train distances}

For the time-resolved distances it is more important when a spike occurs in relation to its neighbours than if there is a match for it in the other spike train. As a consequence timing is more important than rate and in the distance rate is seen through differences in ISI-lengths rather than the spike count. Thus, these measures consider spike timing first and rate differences enter only as a consequence of that timing.

The three time-resolved measures used in this study are the ISI-distance, the SPIKE-distance and the RI-SPIKE-distance.
These measures are calculated by integrating instantaneous dissimilarity values over continuous time rather than summing over discrete spikes. All of these distances are time scale independent since they do not require a time scale parameter.

The time-resolved profiles used to calculate these distances are obtained by comparing the spike trains based on previous and following spikes of both spike trains at each time moment. Due to missing information before the first and after the last spike, edge effect corrections are applied and special cases of empty spike trains and one spike are treated separately by adding auxiliary spikes (See Appendix of \citeauthor{Satuvuori17}, \citeyear{Satuvuori17}).
 
\subsubsection{ISI-distance}

The ISI-distance \citep{Kreuz07c,Kreuz09} assesses the dissimilarity of the spike trains based on instantaneous rate synchrony, where the rate is estimated from the inverse of the local interspike intervals (ISIs). It is calculated by averaging the local rate dissimilarity over the total length of the recording. The ISI-distance obtains the minimum value zero for identical local rates everywhere which means not only perfectly identical spike trains but also spike trains consisting of constant and equal interspike intervals with a global phase shift. It can grow arbitrarily close to one for periodic spike trains with ever larger rate differences.

\subsubsection{SPIKE-distance}

While the ISI-distance assesses local rate dissimilarity, the SPIKE-distance \citep{Kreuz11,Kreuz13} additionally takes into account difference in spike timing. For the simple case of spike trains with steady rates this means that whereas the ISI distance will assess any two processes with the same rate as identical, whereas the SPIKE-distance also evaluates the phase shift. The SPIKE-distance obtains its minimum value zero for exactly identical spike trains only. The theoretical upper limit is one (since this is the limit for the time profile), but in practice the maximum value is 0.55 due to how spikes can be arranged to be as far from each other as possible in the spike trains (as can be shown using, e.g., an evolutionary algorithm).
  
\subsubsection{RI-SPIKE-distance}

The latest time-resolved measure is called rate independent (RI)-SPIKE-distance \citep{Satuvuori17}. While both the ISI-distance and the SPIKE-distance are sensitive to differences in rate, in the RI-SPIKE-distance this sensitivity has been removed. This allows the measure to purely focus on spike timing information.

\subsection{\label{ss:Methods-Analysis}Analysis and statistical considerations}

In this study we use the definition of time coding as correlations beyond rate \citep{Theunissen95} to investigate how the sensitivity of the different spike train distances to rate and time coding depends on the rate of the spike trains. To address this question we follow common practice (see e.g. \citeauthor{Softky93}, \citeyear{Softky93}; \citeauthor{Hanes95}, \citeyear{Hanes95}; \citeauthor{Berry98}, \citeyear{Berry98}) and use pairs of independent steady rate Poisson spike trains as surrogates for random spike trains with fixed rate and no timing information. We sample the distances over multiple realizations in order to calculate the expectation values and to estimate the statistical significance. Any spike trains from spike generating processes that are more similar than expected can be considered containing information beyond pure randomly distributed spikes.  The \highlight{results consist} of two parts, one where we deal with reasonably high overall spike rates and one where we look at the special case of very low rates.

In the first part (Sec. \ref{ss:Detecting}) we reduce the dimension by keeping the sum of the two rates fixed and analyze how the rate ratio determines the expectation value, i.e., the distance value that results of rate coding alone. We then assess the ability of the spike train distances to detect timing information in the data via their operational ranges for temporal coding which we define as the range of values the distances can obtain after differences in rate have already been taken into account.

In the second part (Sec. \ref{ss:Floor}) we examine the floor effect which can occur for spike trains with very low rates. Rate can be estimated by averaging over large numbers of redundant cells or by averaging over repeated presentations of the same stimulus \citep{Bialek90}. The rate is by definition an average quantity and not a property of a single spike train. While counting spikes will always provide an estimate of the rate, the question of similarity is more complicated for spike train distances, since they also contain timing information. 
Thus, first we compare in very general terms, using simple spike counting, estimates of the total rate of two spike train pairs, one pair with equal rates and one pair with a very high rate ratio. In order to have any timing information that can be compared in the data, there must be at least one spike in both spike trains. We again use Poisson models to identify for both spike train pairs the lowest rate at which this is violated with only 5\% probability. In the next step we evaluate for both kinds of spike train distances the full dependence of the distance values on the two rates of the independent Poisson processes. Once more we use a 5\% confidence layer as indicator of a value being outside of the distribution obtained for pure rate.

In both parts of this study we work with spike trains of unit length $1s$, since this way the rate exactly equals the expectation value for the number of spikes in one spike train.
For the spike-resolved Victor-Purpura and van Rossum distance we also look at the influence of the respective time scale parameter.
All our results for both the spike-resolved and the time-resolved spike train distances are gathered in Table \ref{Tbl:All_distances}.

   \begin{figure*}
   	\includegraphics[width = 0.9\textwidth]{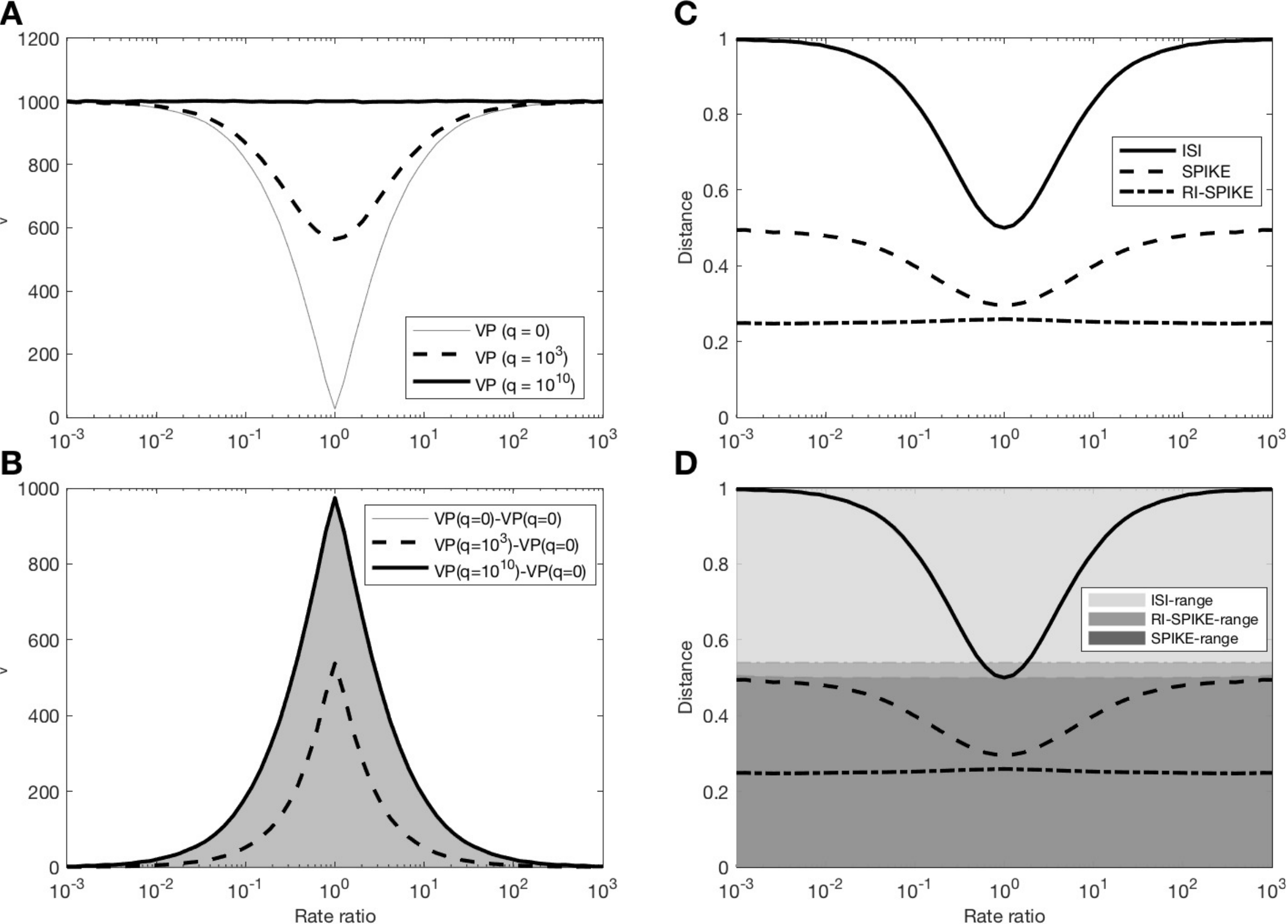}
   	\caption[Figure]{Statistics of two Poisson spike trains with a fixed total rate of 1000Hz divided among the two spike trains. (A) Dependence of the Victor-Purpura distance on the rate ratio 
   	for three different values of the time scale parameter $q$. (B) \highlight{The operational range for spike time coding is marked in grey.} (C) Same dependence for the ISI-distance, the SPIKE distance (compare \citeauthor{Mulansky15}, \citeyear{Mulansky15}) and the RI-SPIKE-distance (see \citeauthor{Satuvuori17}, \citeyear{Satuvuori17}). 
   	(D) Operational ranges for spike time coding. For all distances and rate ratios values can go down to almost zero which is indicated by increasingly darker shades of grey.
   		\label{Fig:V-P_2D}}
   \end{figure*} 
   
\section{\label{s:Results}Results}

In the first part (Sec. \ref{ss:Detecting}) we investigate the rate-dependent sensitivity of the different distances to temporal coding. 
Whereas this analysis is carried out for reasonably high rates, in the second part (Sec. \ref{ss:Floor}) we focus on very low spike rates for which the floor effect can occur.
Each time we first discuss the spike-resolved and then the time-resolved spike train distances.
In the third part we illustrate with simple examples the most important differences between the two kinds of spike train distances.

\subsection{\label{ss:Detecting}Detecting time coding}

Identifying time coding can be difficult, since many spike train distance measures respond to changes in both rate and timing. In order to detect time coding one has to know what is the expected distance when only rate is considered.
%Since any rates contain ‘accidental’ timing information it needs to be taken out before timing correlations can be detected. 
To this aim, we first calculate the expectation values for pure rate code using Poisson spike trains of varying rate ratios by averaging over a sufficient number of realizations (here we use 1000) such that all timing fluctuations cancel out.
\highlight{For the spike-resolved Victor-Purpura distance it is straightforward to eliminate the effect of the rate coding. Here the operational range for time coding, i.e., the range of values that can be obtained beyond rate difference, is easily obtained by subtracting the spike count difference (Victor-Purpura distance for $q = 0$). For the time-resolved distances this is not possible and the operational range becomes simply the range of values.}

% Once we know what is to be expected for spike trains with certain rates, we label the
% values from the expectation value to the minimum that can be achieved by timing as
% operational range of time coding. The larger that range the higher the sensitivity to
% time coding. 
% Note that this operational range of time coding is not the same as simply the range of
% values that can be obtained for the different spike train distances.

\subsubsection{Spike-resolved spike train distances}

\highlight{First we estimate the operational range for time coding for the spike-resolved distances. As a first step} in Fig. \ref{Fig:V-P_2D}A we show the dependence of the Victor-Purpura distance on the rate ratio for three different values of the time scale parameter $q$. For large values of the parameter value q the distance always attains its maximum value independent of the rate ratio. Since the curves are made such that the sum of the rate of the two processes is 1000Hz for a pair of spike trains of unit length 1s, the result is a constant line at $D_V = n_1 + n_2 = 1000$ (compare Section \ref{sss:VP}).
The total rate is kept reasonably high in order not to run into the floor effect.
The only region where the parameter $q$ has any reasonable effect is within the first decade around a rate ratio of one.

In Fig. \ref{Fig:V-P_2D}B we depict the operational range for timing information obtained by subtracting the spike count difference ($q = 0$) which is always part of the total distance value independently of the time-scale parameter. We can see that while for equal rate processes timing information covers almost the whole range of the distance, the operational range is lost very fast with increasing rate difference. As a result, ever larger portions of any distance value come from spike count difference. This holds true even for very large $q$-values which supposedly indicate timing information. 

Thus, in order to obtain spike timing information using the Victor-Purpura distance, the spike trains must have very similar rates. As a consequence of how the distance is defined, the minimum distance for any spike train pair is obtained for $q = 0$. For this parameter value the distance equals exactly the spike count difference between the spike trains. In contrast, when $q$ approaches infinity, the distance becomes spike count over both spike trains and considers spike trains with a smaller overall number of spikes as more similar. 
The only region where the distance has any time coding detection capability is when rates are almost identical and $q$ is in some intermediate range.
While the parameter $q$ is often taken as deciding the relative importance of rate and time coding \citep{Victor05}, this is not the whole story.
We refrain from examining the van Rossum distance in detail, because it behaves similarly to the Victor-Purpura distance but requires normalization between parameter values since the maximum distance value depends on the choice of tau.

Due to these findings we suggest that one should not use the Victor-Purpura distance nor the van Rossum distance for detecting timing information for spike trains that do not have nearly identical rates.
Additionally, since the $q = 0$ case for the Victor-Purpura distance is simply spike count and estimates purely rate we regard it as well suited for detecting rate coding.
These findings are marked at the top of the first main column of Table \ref{Tbl:All_distances} for the normal case of reasonably high rates.

\subsubsection{Time-resolved spike train distances}

In Fig. \ref{Fig:V-P_2D}C we show the rate ratio dependence for the ISI-distance, the SPIKE-distance and the RI-SPIKE-distance. \highlight{The ISI-distance and the SPIKE-distance consider not only timing but also rate information. Accordingly, when only rate is considered, they obtain their minimum value for pairs of spike trains with equal rate. In contrast, the RI-SPIKE-distance ignores differences in rate and thus has a flat response independent of the rate ratio.}

\highlight{Fig. \ref{Fig:V-P_2D}D depicts the operational ranges of all three measures. For any given rate ratio the operational range is defined as the overall range of values from minimum to maximum, i.e. the range that is covered due to deviations from the expectation value caused by the influence of spike timing. All three distances can obtain minimum values arbitrarily close to zero for any rate ratio (an extreme example for large rate ratios would be a infinitesimally narrow multi-spike burst matching a single spike). Regarding the maximum values, the ISI-distance is able to cover the whole interval by approaching the value of one arbitrarily close (the higher rate spike train is evenly distributed, the lower rate spike train is concentrated on the edges). This is not the case for the SPIKE-distance and the RI-SPIKE distance which yield the maximum values of $0.5$ and $0.54$, respectively, for spike train pairs with alternating spikes and the excess spikes of the higher rate spike train concentrated at the edges.} \footnote{\highlight{Note that the maximum values for the SPIKE- and the RI-SPIKE-distance shown here were obtained for sufficiently high spike numbers also in the lower rate spike train. Otherwise the edge effect can lead to slightly higher values (e.g. up to 0.61 instead of 0.5 for the SPIKE-distance). All results for maximum and minimum values were obtained and/or confirmed with an evolutionary algorithm (see, e.g. \citeauthor{Price06}, \citeyear{Price06}).}}

\highlight{The important result here is that, in contrast to the Victor-Purpura distance, all three distances cover a wide range of values even for very high rate ratios. For the ISI-distance and the SPIKE-distance this full operational range means that they are always able to identify both local rate and local spike timing information at the same time. This makes them the best candidates for evaluating synchrony in general. The RI-SPIKE-distance, on the other hand, was specifically designed to ignore differences in rate, so it is obvious that it should not be used for detecting rate coding. However, this very property makes it best suited for detecting pure timing information.}
 
These results are marked in Table \ref{Tbl:All_distances} at the bottom of the first main column for the standard case of reasonably high rates.

%The ISI-distance and the SPIKE-distance are time-resolved distances, and as long as there are enough spikes they can cover a wide range of values for any rate ratio based on both rate and timing information. This allows the distances to consider local rate and spike timing synchrony information at the same time, which leads to a full operational range for all possible rate ratios. 

%While the RI-SPIKE-distance can also obtain almost any value for any rate ratio, it is the only distance that has a flat response to rate differences while maintaining a sufficiently large operational range for timing information. 

%Since the ISI-distance and the SPIKE-distance respond to both rate and time and exhibit a high operational range for differences in spike timing, they are the best candidates for detecting synchrony in general. By synchrony we mean that the distance is able to identify time local differences in both rate and timing regimes.
%The RI-SPIKE-distance, on the other hand, was specifically designed to ignore differences in rate, so it is obvious that it should not be used for detecting rate coding. However, this very property makes it best suited for detecting timing information.

\begin{figure}
	\centering
	\includegraphics[width = 0.48\textwidth]{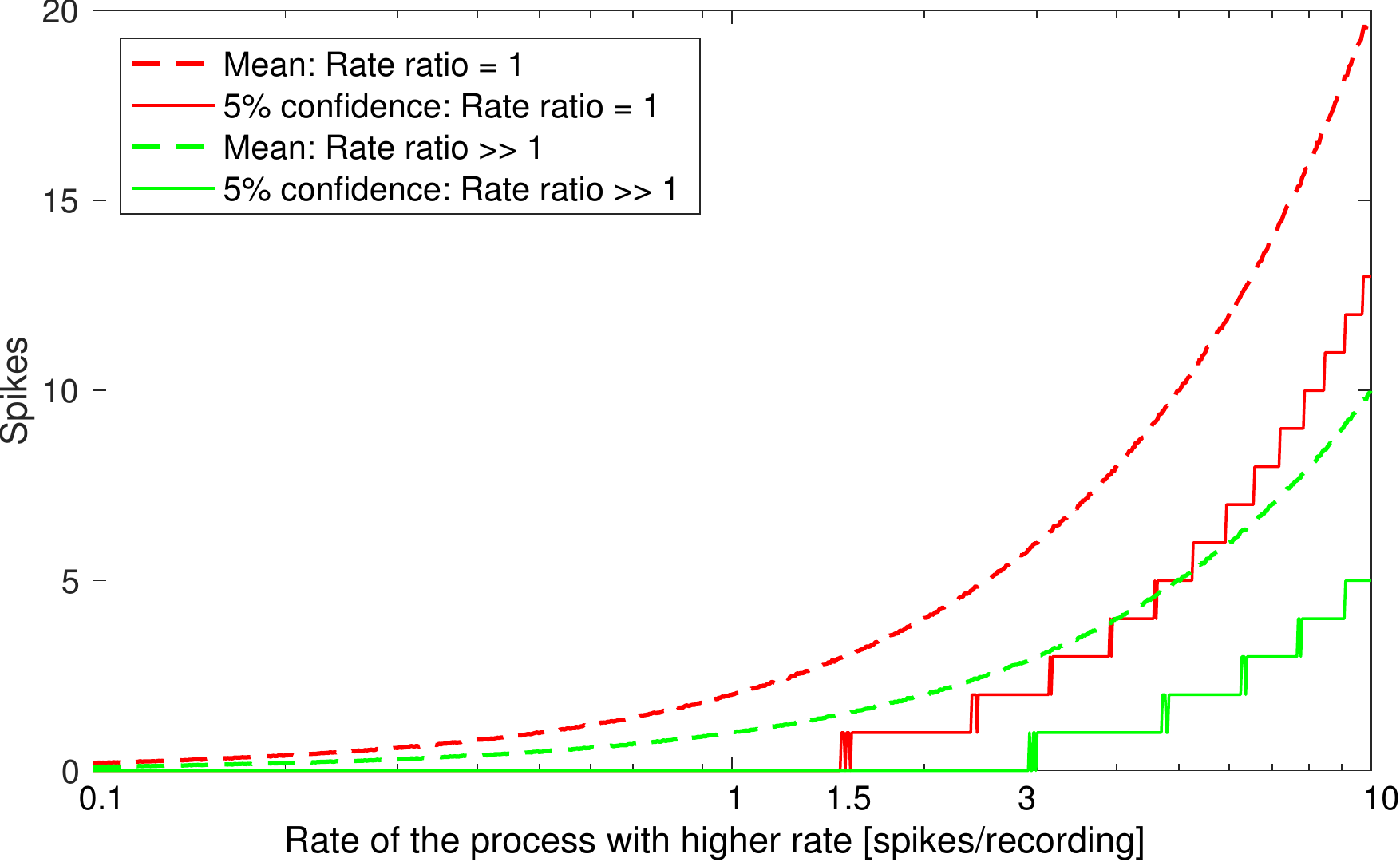}
	\caption[Figure]{The floor effect for spike trains with a very low rate. Mean overall rates (dashed lines) and overall numbers of spikes in the two spike trains at 5\% confidence (solid lines) over 10,000 realizations for a spike train pair with a rate ratio of one (red curves) and a pair with a very high rate ratio (green curves). While the mean values which perfectly match the expectation values (curves not visible) are growing linearly with the rate of the process with higher rate (note the logarithmic x-scale), the actual spike counts can only attain discrete values. Moreover, due to the floor effect it takes \highlight{1.5 spikes} for spike train pairs with rate ratio one and \highlight{3 spikes} for spike train pairs with very high rate ratio to get at least a single spike in either of the two spike trains at 95\% probability.
	
	% ##### Old version: Moreover, due to the floor effect it takes a rate of 1.5Hz for spike train pairs with rate ratio one and a rate of 3Hz for spike train pairs with very high rate ratio to get at least a single spike in either of the two spike trains at 95\% probability. #####
		\label{Fig:Floor}}
\end{figure}

\begin{figure*}
	\includegraphics[width = 0.9\textwidth]{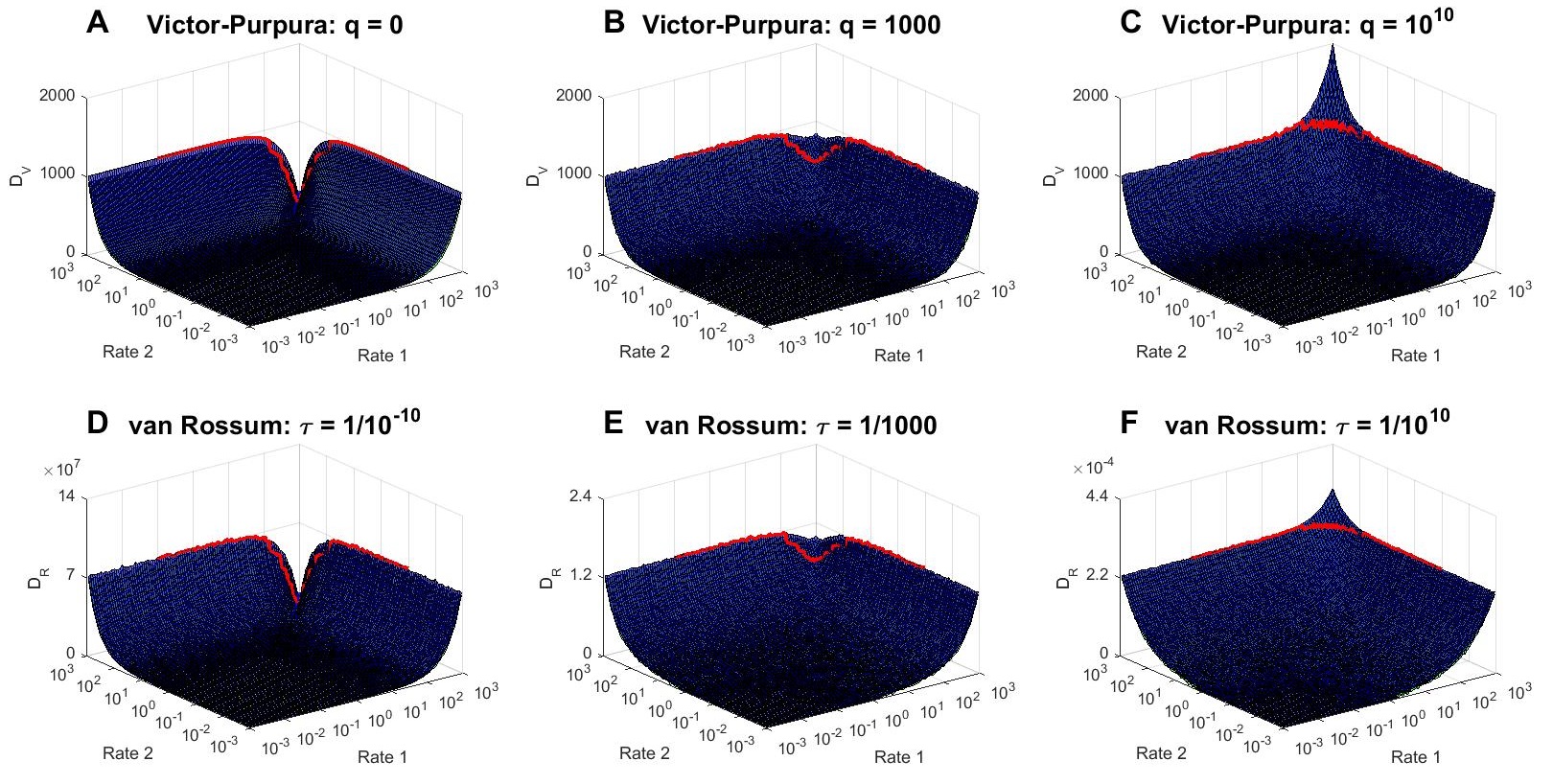}
	\caption[Figure]{Statistical assessment of the Victor-Purpura distance and the van Rossum distance for three different values of the time scale parameter. The blue surface represents the mean of 1000 spike train pair realizations with corresponding rates. The green 5\% confidence boundary trails the mean surface so close that it can not be seen. The red lines show fixed rate producing on average 1000 spikes in total in the two spike trains and thus corresponds to the curves shown in Fig. \ref{Fig:V-P_2D}A. The parameter values for the Victor-Purpura distance are (A) $q = 0$, (B) $q = 1000$ and (C) $q = 10^{10}$. For the van Rossum distance we use the parameter values $\tau = 1/q$. Only for (D) we avoid division by zero by setting $\tau = 1/10^{-10}$. The other two cases yield (E) $\tau = 1/1000$ and (F) $\tau = 1/10^{10}$.
Note that for the Victor-Purpura distance the range of values is only determined by the number of spikes and thus independent of the parameter value. For the van Rossum distance the time scale dependent kernel size changes the range of the distance axis as well. 
\label{Fig:V-P_van_Rossum_3D}}
\end{figure*}

\subsection{\label{ss:Floor}Floor effect}

Even though the rate is a continuous property that can assume any value, the spike trains are sampling the rate with discrete samples. Thus, while for reasonably high rates the "resolution" of the sampling is high enough, as there are sufficient discrete samples available to give a good estimate of the actual rate, at the low end the sampling is not sufficient, since there are only a few discrete values the spike count can assume. This is called the floor effect. 

In order to study the floor effect we first generate 10,000 realizations of two pairs of unit Poisson spike trains, one pair with equal rates (rate ratio = 1) and one pair for which the rate of one of the spike trains is lowered considerably (rate ratio $\gg$ 1). Subsequently, for each of these pairs we calculate the mean and the 5\% confidence boundary from the distribution of total spike counts of the sum of the two trains. This means that at any given rate there is only a 5\% chance of getting spike trains with less spikes than this boundary.

As can be seen from Fig. \ref{Fig:Floor}, the average values follow the actual rate linearly, but the 5\% confidence boundary is discrete, since spike trains can only have an integer number of spikes. Due to this floor effect, most of the rate information is not contained within one short spike train. One needs to have a good sampling of the process either from multiple repetitions or from longer recordings in order to get a meaningful rate estimate. While the mean of the rate is exactly the true rate of the processes, one can have empty spike trains at over 5\% chance when the corresponding confidence curve is at 0. We can see from Fig. \ref{Fig:Floor} that in order to reach non-empty spike trains at 95\% confidence, one should obtain spike trains with an average rate of 3 spikes for high rate ratios and 1.5 for equal rates. This means that if one knows that the two spike trains are from the same process, they need to have a minimum rate of 1.5 spikes/recording. Note that this factor 2 between the 5\% boundary for a rate ratio of 1 and the 5\% boundary for a high rate ratio is observed over all rates. The reason is that for very high rate ratios the spike train with the lower rate tends to be empty and thus non zero samples are only drawn from one spike train instead of two.

\subsubsection{\label{ss:Spike}Spike-resolved spike train distances}

The rate dependence surface is plotted for many different rate combinations and three different parameter values for the Victor-Purpura distance in Fig. \ref{Fig:V-P_van_Rossum_3D}A-C, and for the corresponding parameter values with the van Rossum distance (Fig. \ref{Fig:V-P_van_Rossum_3D}D-F). The first thing to note is that the general shape of the surfaces is very alike between the two spike-resolved distances using the parameter conversion $\tau = 1/q$. In the following we discuss the Victor-Purpura distance in more detail, since the definition for it is more intuitive.

As can be seen in Fig. \ref{Fig:V-P_van_Rossum_3D}A-C, the smooth mean surfaces trace exactly the total rate of the processes for high rate ratios identical to Fig. \ref{Fig:Floor} mean spike counts and the 5\% confidence layer traces so close by to the mean that they cannot be distinguished. 
It is important to note that there is no visible artefact from the floor effect in the distance measure but the curves are smooth, since the distance is always primarily spike count difference.
This is caused by the property that the parameter $q$ only has an effect in a very narrow area near identical rates as we have already seen in Sec. \ref{ss:Detecting}. For low rates the Victor-Purpura distance is almost flat and close to zero since the range $[n_2-n_1, n_2+n_1]$ tends to vanish.

In Fig. \ref{Fig:V-P_van_Rossum_3D}D-F we can see that for the van Rossum parameter $\tau = 1/q$ the distance performs very similar to the Victor-Purpura distance. However, a comparison of distances obtained with different parameter values can be done with the Victor-Purpura distance, but not with the van Rossum distance since there the range of values also depends on the time scale parameter. 

Both distances can be used for very low rates without artefacts, since they assess rate first and timing second. These findings are marked at the top of the second main column of Table \ref{Tbl:All_distances}.

\begin{figure*}
	\includegraphics[width = 0.9\textwidth]{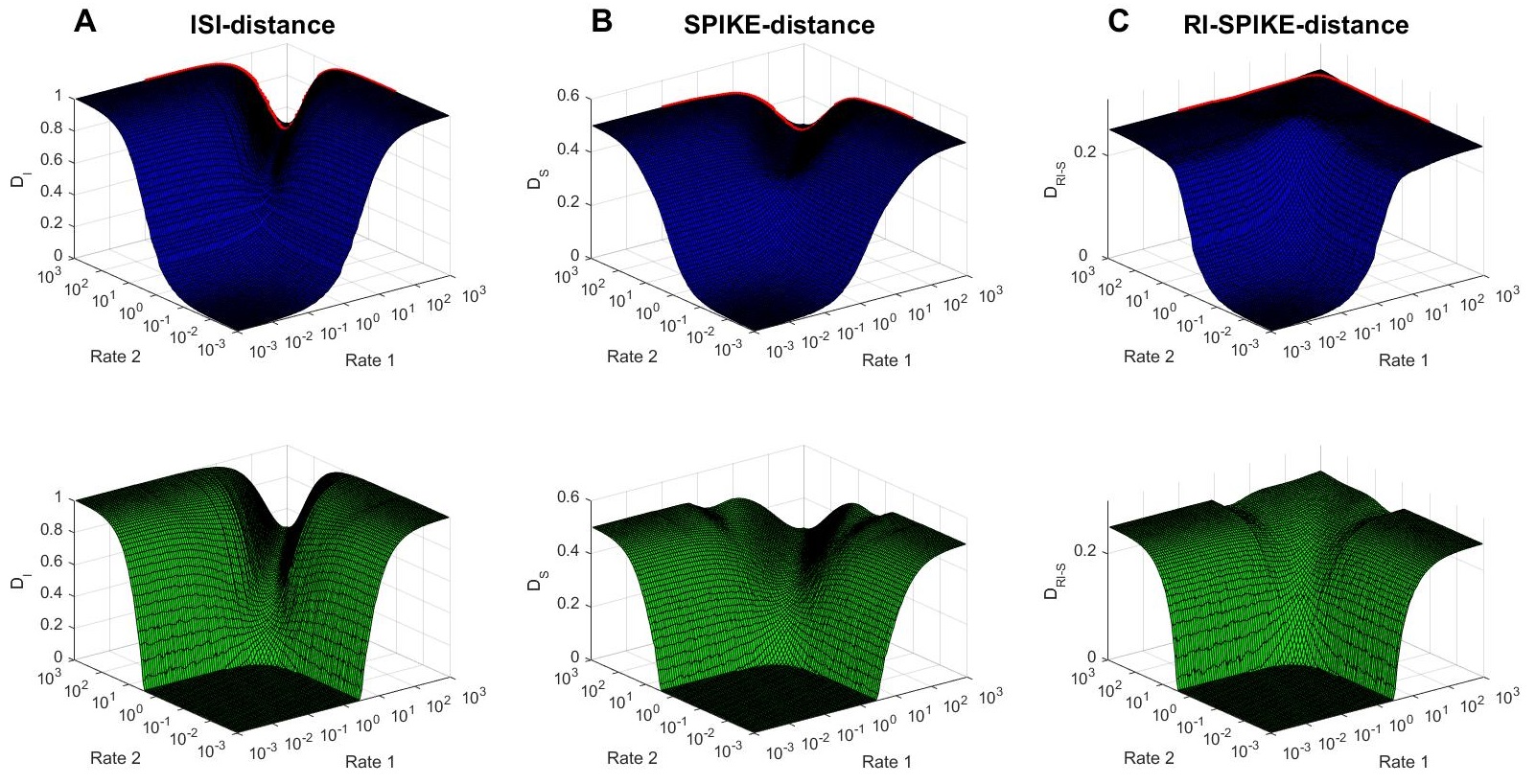}
	\caption[Figure]{Same as Fig. \ref{Fig:V-P_van_Rossum_3D}, but this time for the ISI-distance (A), the SPIKE-distance (B) and the RI-SPIKE-distance (C). For the sake of visibility we separate the mean vales (blue, top) and the 5\% confidence boundary (green, bottom) in two separate subplots. The red line represents the fixed rate equivalent to the curves in Fig. \ref{Fig:V-P_2D}C. 
		\label{Fig:ISI_SPIKE_3D}}
\end{figure*}

\subsubsection{\label{ss:Time}Time-resolved spike train distances}

In Fig. \ref{Fig:ISI_SPIKE_3D} are again plotted the mean and 5\% confidence boundary for the distribution, but this time for the ISI-distance, the SPIKE-distance and the RI-SPIKE-distance. The means of the distance values grow once the random processes start introducing spikes. In Fig. \ref{Fig:ISI_SPIKE_3D}A one can see the mean rate surface of the ISI-distance.  While the surface increases almost linearly up to 1Hz, the scaling to total time makes the distance settle to an expectation value that depends on the rate ratio. However, it is important to note, that the 5\% confidence layer does only get above zero after a threshold rate is reached. This means that below this threshold there is at least a 0.05 chance of getting two empty spike trains due to the floor effect (compare again Fig. \ref{Fig:Floor}).

In Fig. \ref{Fig:ISI_SPIKE_3D}B we can see that the general shape of the mean surface for the SPIKE-distance shares the rate-dependent nature of the ISI-distance. On the other hand, the RI-SPIKE-distance (Fig. \ref{Fig:ISI_SPIKE_3D}C) by construction shows no rate dependence. This means that the distance is purely based on spike timing and ignores rate differences in the spike trains.

In contrast to the spike-resolved Victor-Purpura and van Rossum distances, the time-resolved ISI-distance, SPIKE-distance and RI-SPIKE-distance can attain any distance value between zero and the maximum even for very low rates. 
While both kinds of distances are affected by the same floor effect, the definition of time coding by \citet{Theunissen95} as being over and above any information coded in rate is not satisfied for time-resolved distances, since the measures attempt using timing information even before there is sufficient rate.
Only once it becomes unlikely to have two empty spike trains the boundary starts to increase and the values obtained start being reliable.
This increase is not only manifested for empty spike trains. Already spike trains from processes with lower rates are more likely to be similar by chance than those obtained with higher rates. From this we can conclude that spike train distance values can only be considered non-random, when the rate of the spike-generating process is high enough not to produce empty (or quasi empty) spike trains and when the similarity of a pair is below the 5\% confidence interval surface. Even for these the similarity has to be very pronounced not to be considered as being drawn from a random distribution. 

The ISI-distance, the SPIKE-distance and the RI-SPIKE-distance are time-resolved which allows an instantaneous assessment of similarity. The normalization of the measures means that the spikes are assessed in relation to the length of the local ISIs. This time-scale independence allows comparisons of spike trains with very different rates.

The reason why the ISI-distance, the SPIKE-distance and the RI-SPIKE-distance can draw any values even for spike trains that have very few if any spikes is an artefact from the normalization used for these measures. Since the values are time-resolved, and even a single spike needs to be comparable, the measures apply edge effect corrections \citep{Satuvuori17}. For this reason already for such extreme cases as a spike train pair composed of an empty spike train and a spike train with just one spike the distances can obtain virtually all of their range (depending on the location of this single spike).
While the time-resolved nature gives the measures the advantage of being able to assess similarity in time, instead of just assessing timings of pairs of spikes, it becomes a downside for very low spike counts when not enough information is available to form meaningful time-resolved profiles. Because of this we do not suggest using time-resolved measures for processes with a very low rate. These results for the ISI-distance, the SPIKE-distance and the RI-SPIKE-distance can be found in Table \ref{Tbl:All_distances} at the bottom of the second main column for very low rates.

\subsection{\label{ss:Examples}Examples}

%\begin{figure*}
%	\includegraphics[width = 0.9\textwidth]{New}
%	\caption[Figure]{Examples.
%		\label{Fig:Examples2}}
%\end{figure*}

\begin{figure*}
	\includegraphics[width = 0.9\textwidth]{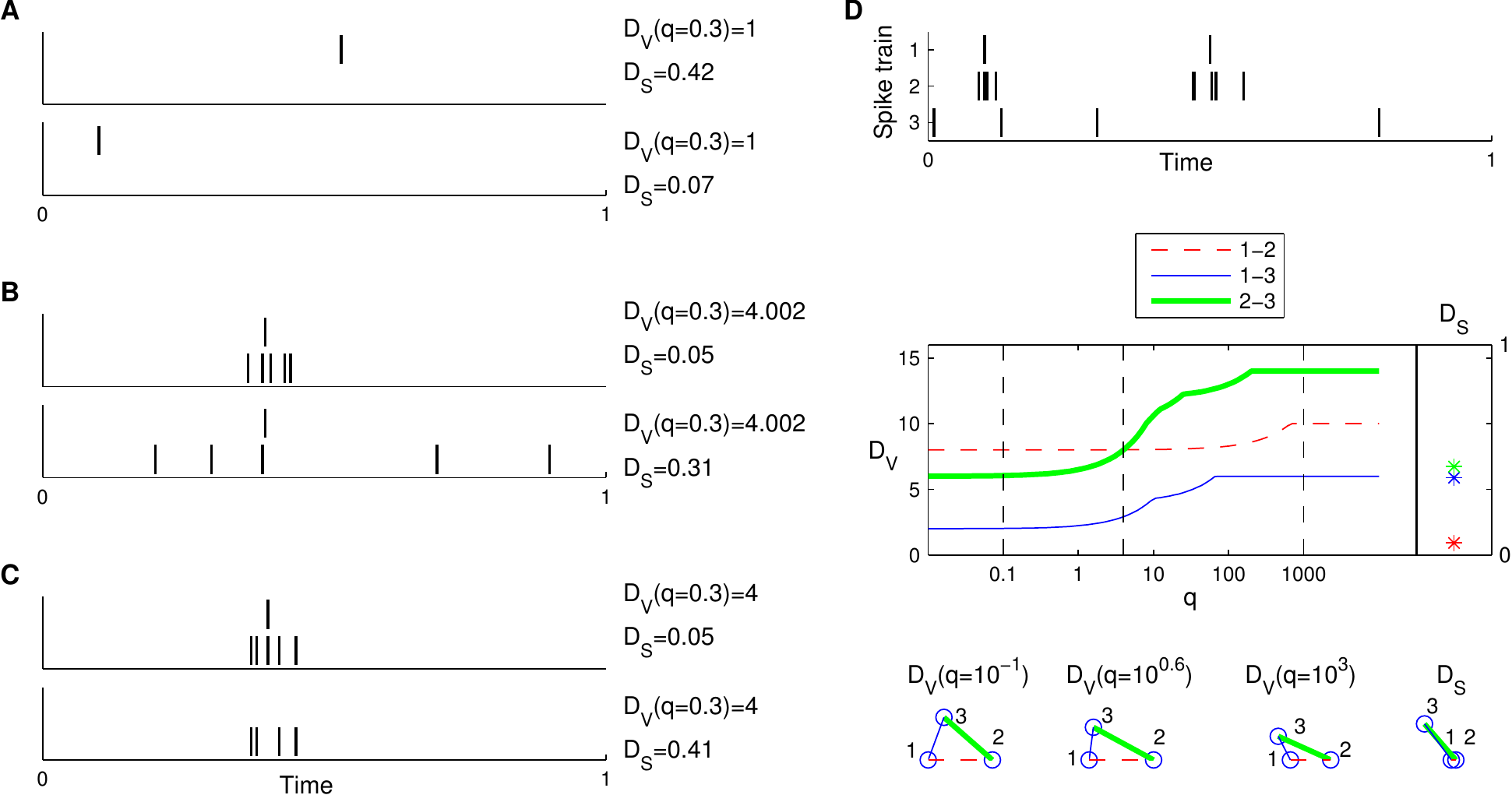}
	\caption[Figure]{Simple examples illustrating some of the most important differences between the \highlight{Victor-Purpura distance $D_V$ and the SPIKE-distance $D_S$}. (A) Floor effect: SPIKE-distance looks for timing in a single spike. (B) Spike trains with different rates: Victor-Purpura distance ignores timing information of extra spikes. (C) The Victor-Purpura distance is insensitive to exactly matching spikes. (D) Simple clustering example: In contrast to the SPIKE-distance, for spike trains with different rates the Victor-Purpura distance can never really focus on the timing information, even for large q-values.
     \label{Fig:Examples}}
\end{figure*}

In previous Sections we have done statistical analysis of the spike train distances. In this Section we give a brief overview of the implications of the results using constructed examples (Fig 5). In Fig \ref{Fig:Examples}A we see typical examples of spike trains near the floor effect. There is hardly any visible rate and thus determining timing is problematic. Since the Victor-Purpura distance is primarily assessing rate and only second single spike timing, it gets the distance of one for both spike train pairs. However, the SPIKE-distance always incorporates timing and thus it can reach a very large variety of different values (0.42 and 0.07 in these example spike trains) even when there is very little timing information. This is due to the inflated contribution of the edge effects.

In Fig. \ref{Fig:Examples}B we see another situation, where we have two spikes at a short distance of each other. It is clearly visible that in the first spike train pair there is more timing information beyond rate than in the second. However, since the Victor-Purpura distance considers only the closest spike pairs, the distance will be the same independently of how the four other spikes are arranged. This is again due to rate first assessment of similarity. The SPIKE-distance uses timing assessment over time rather than over spikes and indicates a clear distinction between the two cases.

However, even the Victor-Purpura distance itself is ambiguous in respect to whether the value is obtained from rate or from timing. In Fig. \ref{Fig:Examples}C we see two very different scenarios. In the top spike train there is a perfectly coincident pair of spikes plus a few additional spikes in the second spike train. If we compare this to the exactly identical spike trains with only the two coinciding spikes removed we get exactly the same distance. In both cases this value comes purely from rate difference. However, there is a considerable difference in timing correlation between the two spike trains. This is clearly reflected by the values obtained for the SPIKE-distance.

In Fig. \ref{Fig:Examples}D we show how serious these effects are. As we mentioned already in Section \ref{ss:Detecting}, the Victor-Purpura distance and the van Rossum distance can hardly distinguish any timing information once the rate ratio is 10-fold or more. However the effects are realized already for much lower rate differences. In this example we consider three spike trains (2, 10 and 4 spikes). The first two spike trains clearly contain timing correlations, while the third does not correlate with either of the two. We can observe that the rate difference obtained with low q values (here $q = 10^{-2}$) is basically the spike count difference between pairs. The result is that the spike trains 1 and 3 have more similar rate than any other pair. Then by tracking the distances for increasing q-values the order of the distances remains the same until single spike timing reaches its peak at $q = 10^{0.6}$. Here the distances from 1 to 2 and 2 to 3 are equal. However, still the distance between 1 and 3 prevails as the smallest one indicating that they are the most similar pair. While after this q-value the distance from 1 to 2 is closer than 2 to 3, the pair with built in time correlations, 1 and 2 never reaches the range where it would be the most similar. In Fig. \ref{Fig:Examples}D relative distances are drawn as distance triangles with the respective distances for different q. For comparison we have added the distance triangle obtained for the spike trains using the SPIKE-distance, which, since it is not restricted to comparing pairs of spikes and thus rate first, finds the most time correlated pair with ease.

\section{\label{s:D&C}Discussion and Conclusions}

% Summary 

% Take home 1: spike vs time resolved
Spike train distances can be constructed in a few different ways. The basic components common to all are the spikes in time.
Some spike train distances are evaluated over values attached to spikes, like the Victor-Purpura distance \citep{Victor96,Victor97}, and the van Rossum distance \citep{VanRossum01}, and thus difference in spike count becomes a dominant feature. We call these spike-resolved distance measures. For the ISI-distance \citep{Kreuz07c,Kreuz09}, the SPIKE-distance \citep{Kreuz11,Kreuz13}, and the RI-SPIKE-distance \citep{Satuvuori17} effects of spikes are evaluated in relation to time and these distances are thus time-resolved.

 % Take home 2: research questions 
In this study we asked two questions: How does the sensitivity of the different spike train distances to rate and time coding depend on the rates of the two processes and how high a rate is needed in order to obtain reliable estimates of timings in the data? To answer these questions we used two independent steady rate Poisson spike trains as surrogates for rate only coding neurons and calculated both the expectation values and the 5\% confidence boundary over multiple realizations. The results are gathered in Table \ref{Tbl:All_distances}. 

% Take home 3: results of spike-resolved
The first key finding of the analysis of time coding is that the spike-resolved Victor-Purpura distance compares the spike trains spike for spike and thus they are always sensitive to differences in spike counts even for parameter values seemingly indicating time coding.
For large spike count differences the spike-resolved distances do not obtain the ability to assess timing information beyond spike pairs and thus in many cases most of the distance comes from mismatch in spike counts rather than timings, independently of the time scale parameter.
As a result, for the Victor-Purpura distance timing information is only available for spike trains with almost identical rates (as illustrated in Fig. \ref{Fig:Examples}B and \ref{Fig:Examples}D). Since the behaviour of the van Rossum distance in response to rate differences closely resembles that of the Victor-Purpura distance, it also has the same problem (in addition to its normalization issues for different tau-values). 
These results are consistent with those obtained by \citet{Chicharro11} for a similar analysis of rate differences.

% Take home 4: results of time-resolved
The second key finding is that the time-resolved measures perform better in assessing timings in the normal case of reasonably high rates. 
Also these measures can provide a meaningful instantaneous similarity profile within the coding window. Since they assess similarity in time, the exact spike count becomes less important and the actual timing of events becomes more relevant.
However, they suffer from artefacts when the rates of the spike generation processes is so low that the floor effect takes place (see Fig. \ref{Fig:Examples}A for an example).
For spike trains with only a few spikes one should use the spike-resolved Victor-Purpura or van Rossum distance, since they assess first similarity in spike count and then apply timing information assessment only for pairs of spikes.

% Discussion

Investigation for neuronal coding has been going on for decades and the two most prominent approaches are to find similarities in responses in rate and in timing. However, while the distinction between the coding types is clear in a philosophical sense as presented by \citet{Theunissen95}, the exact nature of time coding as being "over and above any information that might be correlated with the number of spikes within the window" is not uniquely defined. It depends on the type of correlation chosen. 

For the spike-resolved distances sensitivity to rate never goes away since the information contained in the timing accuracy of spike pairs is always added on top of the rate information. For the Victor-Purpura distance, even for high q-values the relative importance of timing gets smaller for increasingly different rates. Moreover, this timing information only includes the distances between spikes that are needed to match all the spikes of the shorter spike train with their nearest spikes in the other spike train. While the importance of these differences does increase with higher q-values (Figure \ref{Fig:V-P_2D}B) it is still capped by the maximum cost of 2 per spike pair (‘delete and add’ instead of ‘shift’). The timing of the other $n_2-n_1$ spikes in the longer spike train is ignored entirely (see Fig. \ref{Fig:Examples}B), a loss of information that again increases with larger rate difference. Additionally, it is ambiguous if the distance obtained with parameter values indicating timing truly come from timing as shown in Fig. \ref{Fig:Examples}C. Therefore, while it may sound the intuitive thing to do, one cannot simply take the distance with a parameter indicating time coding and subtract the rate coding distance to get the timing correlations in the data.
For the time-resolved ISI-distance, SPIKE-distance and RI-SPIKE-distance the measures have been defined as the integral over a dissimilarity profile that covers the whole recording time. In this case the intuitive difference between coinciding burst and steady rate is accounted for, since the assessment is not done spike by spike. 
On the other hand, for the ISI-distance or the SPIKE-distance there is no single value to be obtained for rate coding in order to subtract the time coding information content without a surrogate.
The only measure we have shown to be independent of rate information is the RI-SPIKE-distance.
This is an important notion, since one of the most essential questions in the analysis of neuronal coding is if spike trains contain information beyond rate and this measure is able to provide exactly this assessment.

For the investigation of neuronal coding the argument between rate and time coding types hinges crucially on the inaccuracy of the definition of time coding. Additionally the coding types are slightly mixed through the concept of the encoding window. If one estimates rate over an encoding window and then splits it into multiple bins, which are essentially shorter encoding windows, the result will be an assessment of timing and timing accuracy depending on the bin size.

% Does the typical $D_V(q)$ analysis still make sense?

Based on our analysis we would advice against using the spike-resolved Victor-Purpura or van Rossum distances if one is interested in timings in the data when the rates are sufficiently high to avoid any floor effect. Also, the original interpretation of the parameters $q$ and $\tau$ as precision of temporal coding (\citeauthor{Victor05}, \citeyear{Victor05}, but see also \citeauthor{Chicharro11}, \citeyear{Chicharro11}) is slightly misleading in the light of this study, since it only works for nearly identical rates. In all cases, the information about differences in rate is always included. This seems to be consistent with results found in \citet{LopesDosSantos15}.
As a result, it might be useful to reassess some older studies, where the Victor-Purpura distance and the van Rossum distance have been used for the distinction of time coding from rate coding in neuronal data such as the studies conducted by \citet{wang07} and \citet{tang14} for songbird data.

If one is interested in rate, the Victor-Purpura distance and the van Rossum distance can provide information of the rate difference between the spike trains. However, this information is equally available from the spike count and the true strength of the measures lies in assessing synchrony of spike trains with a very low rate, where they can provide a distance at the same time based on rate difference and on timings of single spikes.
The results for the very low rate can be relevant in real data analysis, but most of the time there are enough spikes to use the time-resolved distances.
The time-resolved distances always assess timing information due to how they are defined.

% Unit spike train and implications

For simplicity the results in this paper have been obtained for spike trains of unit length 1s, since for a spike train of length one any rate will produce on average the same number of spikes as the rate. However, the values will scale with the recording length. This is very important, since a rate requirement of 1.5Hz for a spike train of unit length 1s will translate to 15Hz for a 100ms recording. Also the analysis performed here can be used to estimate window or bin sizes for methods that need to split a spike train into smaller segments. For this, assuming a steady rate over a recording, one can obtain a very good rate estimate via a better sampling of the process, since the original non-divided spike train is less likely to suffer from the floor effect.

It is important to note that from a statistical point of view the rate of the process and the length of the recording are inversely proportional.
Recording a 10Hz process for 0.1s gives exactly the same amount of information about the distribution of the process as recording a 1Hz process for 1s. Thus it is possible to obtain the statistical significance of the rate of a steady rate process either by taking one long recording or averaging over multiple short ones. Taking multiple short samples of a process may be more laborious than using one longer recording. On the other hand, it is harder to ensure stationary of the process over longer time than over multiple repetitions. Another problem with neurons is to ensure that the rate of the process does not change faster than it is sampled. However, assessing change in the rate of the process in relation to sampling is outside the scope of this paper.

% Poisson process as random / Outlook

In this study random spike trains were simulated as steady rate Poisson processes. While this approach is often used, it does not match many experimental ISI-distributions (e.g. \citeauthor{Softky93}, \citeyear{Softky93}; \citeauthor{Baddeley97}, \citeyear{Baddeley97}). 
It is important to use a meaningful surrogate when evaluating whether the distance could have been obtained by chance.
While there have been some studies on spike train surrogates \citep{Rapp94b,Schreiber00,Louis10}, there is yet no simple answer as to which null hypothesis to test for and how to adapt the surrogates to the specific null hypothesis that is tested. 

% Outlook

Here we compared four established spike train distance measures and one recently proposed measure. The same study could be conducted on other measures, e.g. on SPIKE-synchronization \citep{Kreuz15} or on other new classes of measures \citep{Rusu14}, to see how these  approaches perform under the same conditions. 
It would also be interesting to see if spike-resolved and time-resolved distances all share some common characteristics. Also different kinds of normalizations and integrations in the measure descriptions may share common features. If one were able to construct a theoretical framework for the distances, perhaps by combining desired properties from complementary measures it could be possible to construct a measure that works universally in all cases. In the meantime, we suggest referring to Table \ref{Tbl:All_distances} when deciding which of these measures to use for which kind of data.  

The implementations for the ISI-distance, the SPIKE-distance and the RI-SPIKE-distance are provided online in three separate freely available code packages called
SPIKY\footnote[1]{http://www.fi.isc.cnr.it/users/thomas.kreuz/Source-Code/SPIKY.html}
(Matlab graphical user interface, \citeauthor{Kreuz15}, \citeyear{Kreuz15}), PySpike\footnote[2]{http://mariomulansky.github.io/PySpike/}
(Python library, \citeauthor{Mulansky16}, \citeyear{Mulansky16}) and, most recently, cSPIKE
\footnote[3]{http://www.fi.isc.cnr.it/users/thomas.kreuz/Source-Code/cSPIKE.html}
(Matlab command line with MEX-files). Source codes for the Victor-Purpura distance and the van Rossum distance are available as well\footnote[4]{http://www.fi.isc.cnr.it/users/thomas.kreuz/sourcecode.html}.

\section*{\label{Acknowledgements} Acknowledgements }
\footnotesize

We thank Irene Malvestio for many useful discussions.
We gratefully acknowledge support from the European Union's Horizon 2020 research
and innovation program under the Marie Sklodowska-Curie Grant Agreement No. 642563
'Complex Oscillatory Systems: Modeling and Analysis' (COSMOS).

\normalsize

\appendix

\section{\label{App:SpikeResolved} Spike-resolved spike train distances}

In this paper we evaluate two spike-resolved distances, the Victor-Purpura distance \citep{Victor96,Victor97} and the van Rossum distance \citep{VanRossum01}.

\subsection{\label{App:V-P} Victor-Purpura distance}
The Victor-Purpura distance $D_V$ \citep{Victor96,Victor97} is calculated by finding the smallest path to convert one spike train into the other using three elementary steps: 

\begin{enumerate}
	\item Deleting a spike with a cost of 1.
	\item Inserting a spike with a cost of 1.
	\item Shifting a spike to coincide with another spike in the other spike train with a cost of $q|\Delta t|$.
\end{enumerate}
The time scale parameter $q$ determines how far away two spikes can be in order for it to cost less than achieving the same by using steps 1 and 2. This parameter is thus considered as an indicator of the relative importance between time and rate coding.

\subsection{\label{App:v-R}van Rossum distance}
The van Rossum distance $D_R$ \citep{VanRossum01} first transforms discrete spikes into continuous functions by convolving each spike with an exponential kernel
\begin{equation} \label{eq:van_R_kernel}	
	H(t)exp( - \frac{t}{ \tau } ),	
\end{equation}
where $H(t)$ is the Heaviside step function, t is time and $\tau$ the time constant. Using the resulting waveforms $\tilde{x}(t)$ and $\tilde{y}(t)$ the distance can then be calculated as
\begin{equation} \label{eq:van_R_distance}	
D_R(\tau) = \frac{1}{ \tau }\int_{0}^{\infty} |\tilde{x}(t)-\tilde{y}(t)|^2 dt.	
\end{equation}
Quite recently a markage trick has been presented which significantly reduces the computational cost of calculating the van Rossum distance \citep{houghton12}.

\section{\label{App:TimeResolved} Time-resolved spike train distances}

In this paper we also investigate three time-resolved spike train distances, the ISI-distance \citep{Kreuz07c,Kreuz09}, the SPIKE-distance \citep{Kreuz11,Kreuz13}, and the very recently proposed RI-SPIKE-distance \citep{Satuvuori17}. Note that in \citet{Satuvuori17} all of these distances have been adapted for data with multiple time scales. To see how these adaptive versions behave please refer to \ref{App:Adaptives}.

\subsection{\label{App:ISI}ISI-distance}

The ISI-distance $D_I$ \citep{Kreuz07c,Kreuz09} measures the instantaneous rate difference between spike trains.
It relies on a time-resolved profile, meaning that a dissimilarity value is defined for each time instant.
To obtain this profile, we assign to each time instant $t$ the time of the previous spike
\begin{equation} \label{eq:Prev-Spike}
t_{\mathrm {P}}^{(n)} (t) = \max\{t_i^{(n)} | t_i^{(n)} \leq t\}  \quad
\textrm{for }\quad t_1^{(n)} \leqslant t \leqslant t_{M_n}^{(n)}
\end{equation}
and the time of the following spike
\begin{equation} \label{eq:Foll-Spike}
t_{\mathrm {F}}^{(n)} (t) = \min\{t_i^{(n)} | t_i^{(n)} > t\} \quad \textrm{for } \quad
t_1^{(n)} \leqslant t \leqslant t_{M_n}^{(n)}.
\end{equation}
From this for each spike train $n$ an instantaneous interspike interval (ISI) can be calculated as
\begin{equation} \label{eq:ISI}
x_{\mathrm{ISI}}^{(n)} (t) = t_{\mathrm {F}}^{(n)} (t) - t_{\mathrm {P}}^{(n)} (t).
\end{equation}
The pairwise ISI-profile is then defined as
\begin{equation} \label{eq:pairwise ISI-profile}
I_{n,m} (t) = \frac{|x_{\mathrm{ISI}}^{(n)} (t) - x_{\mathrm{ISI}}^{(m)}
	(t)|}{\max \{x_{\mathrm{ISI}}^{(n)} (t), x_{\mathrm{ISI}}^{(m)} (t)\} }.
\end{equation}
The multivariate ISI-profile is obtained by averaging over all pairwise ISI-profiles:
\begin{equation} \label{eq:multivariate}
I (t) = \frac{2}{N(N-1)}\sum_{n=1}^{N-1} \sum_{m=n+1}^N I_{n,m} (t).
\end{equation}
Finally, integration over time gives the distance value
\begin{equation} \label{eq:distance}
D_I = \frac{1}{t_e-t_s} \int_{t_s}^{t_e} I (t) dt .
\end{equation}
Here, $t_s$ and $t_e$ denote the start and the end of the recording, respectively.

\subsection{\label{App:SPIKE}SPIKE-distance}

The SPIKE-distance $D_S$ \citep{Kreuz11,Kreuz13} measures the relative spike timing between spike trains normalized to local firing rates.
In order to assess the accuracy of spike events, each spike is assigned the distance to its
nearest neighbor in the other spike train
\begin{equation} \label{eq:Closest spike}
\Delta t_i^{(n)} = \min_j(|t_i^{(n)} - t_j^{(m)}|).
\end{equation}
These distances are then interpolated between spikes using for all times $t$ the time differences
to the previous spike
\begin{equation} \label{eq:Weighting_prev}
x_P^{(n)}(t) = t-t_i^{(n)}    \quad \textrm{for }\quad  t_i^{(n)} \leqslant t \leqslant
t_{i+1}^{(n)},
\end{equation}
and to the following spike
\begin{equation} \label{eq:Weightings_foll}
x_F^{(n)}(t) = t_{i+1}^{(n)}-t \quad \textrm{for }\quad  t_i^{(n)} \leqslant t \leqslant
t_{i+1}^{(n)}.
\end{equation}
This defines a time-resolved dissimilarity profile from discrete values the
same way as Eqs. \ref{eq:Prev-Spike} and \ref{eq:Foll-Spike} did for the ISI-distance.
The instantaneous weighted spike time difference for a spike train can then be calculated
as the interpolation from one difference to the next
\begin{equation} \label{eq:weighted distance}
S_{n} (t) = \frac{\Delta t_i^{(n)} (t)x_F^{(n)}(t) + \Delta t_{i+1}^{(n)}
	(t)x_P^{(n)}(t)}{x_{\mathrm{ISI}}^{(n)} (t)}\quad,\quad  t_i^{(n)} \leqslant t \leqslant
	t_{i+1}^{(n)}.
\end{equation}
This function is analogous to the term ${x_{\mathrm{ISI}}^{(n)}}$ for the ISI\hyp{}distance,
with the only difference that it is piecewise linear instead of piecewise constant.
It is also continuous.

The pairwise SPIKE-distance profile is then obtained by averaging the weighted spike time differences,
normalizing to the local firing rate average and, finally, weighting each profile by the
instantaneous firing rates of the two spike trains
\begin{equation} \label{eq:SPIKE profile}
S_{m,n} (t) = \frac{S_n x_{\mathrm{ISI}}^m (t) +S_m x_{\mathrm{ISI}}^n (t) }
	{2\langle x_{\mathrm{ISI}}^{n,m} (t)\rangle^2 }.
\end{equation}
From this the multivariate profile and the distance value can be calculated similar to Eqs. \ref{eq:multivariate} and \ref{eq:distance}.

\subsection{\label{App:RI-SPIKE}RI-SPIKE-distance}

The rate-independent SPIKE\hyp{}distance (RI-SPIKE-distance) $D_S^{RI}$ \citep{Satuvuori17} is similar to the SPIKE-distance, but leaves out the weighting by rate difference by substituting Eq. \ref{eq:SPIKE profile} with
\begin{equation} \label{eq:RI-SPIKE profile}
S_{m,n}^{RI}(t) = \frac{S_n(t) + S_m(t)} {2\langle x_{\mathrm{ISI}}^{n,m} (t)\rangle}.
\end{equation}
The RI-SPIKE-distance shares all the properties of the SPIKE-distance, but it only evaluates
normalized spike timing differences, whereas the SPIKE-distance additionally uses differences
in rate to determine similarity.

Again, the multivariate profile and the distance value can be obtained analogous to Eqs. \ref{eq:multivariate} and \ref{eq:distance}.

 \begin{figure}
 	\includegraphics[width = 0.49\textwidth]{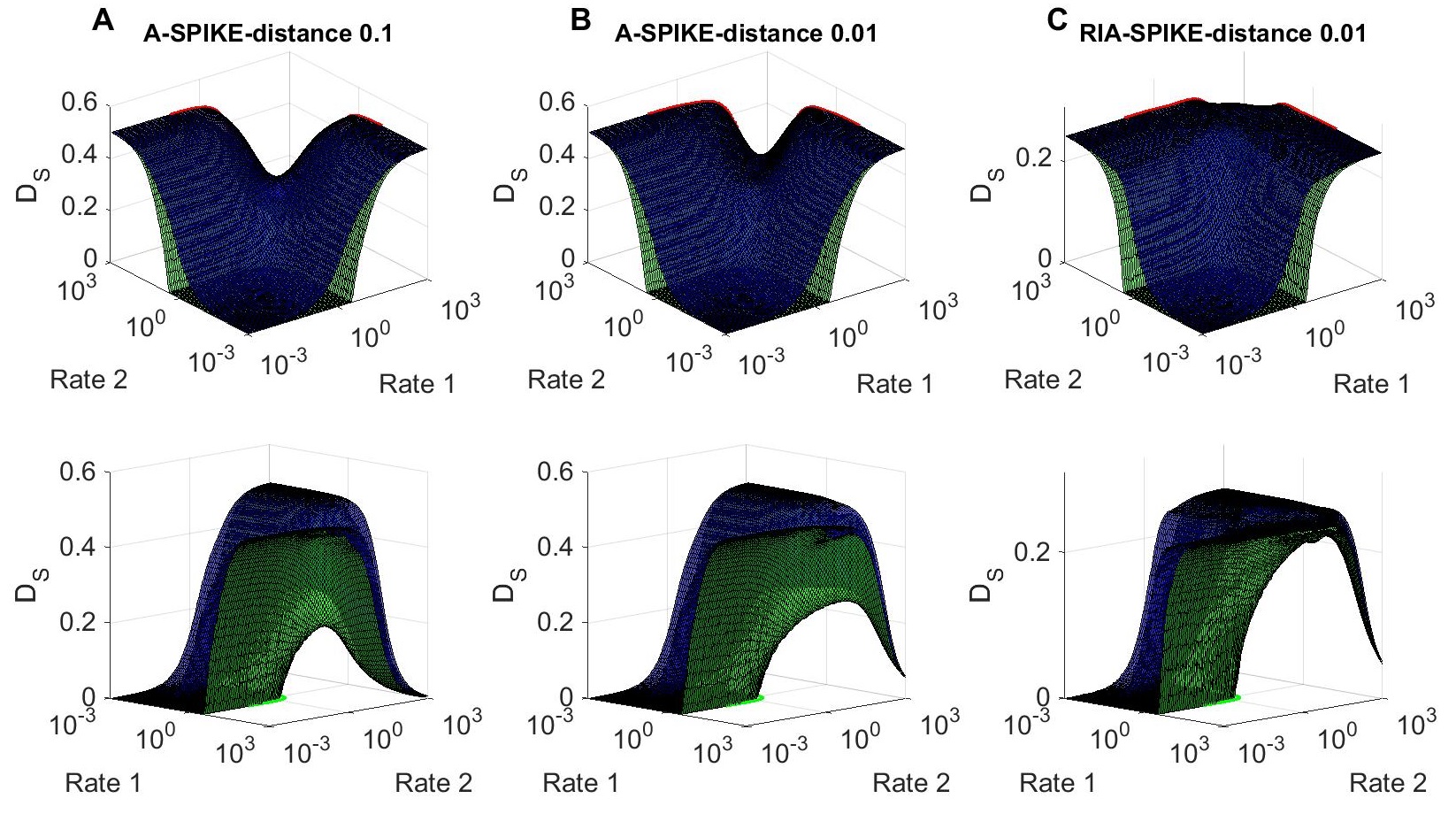}
 	\caption[Figure]{Adaptive spike train distances A-SPIKE-distance and RIA-SPIKE-distance with fixed parameters $\mathcal{T}$. In each case we present two different surface projections. (A) A-SPIKE-distance with $\mathcal{T} = 0.1$ applied to two Poisson spike trains of unit length. When both spike trains exhibit high rates compared to the global parameter $\mathcal{T}$ the spike trains are considered more similar. (B) A-SPIKE-distance with $\mathcal{T} = 0.01$. With smaller details still considered important the drop moves to higher frequencies. (C) RIA-SPIKE distance with $\mathcal{T} = 0.01$. The RIA-SPIKE distance does not take into account rate, but starts ignoring differences in spike times once both rates become high. 
 		\label{Fig:A_ISI_SPIKE_3D}}
 \end{figure}

\subsection{\label{App:Adaptives}Adaptive spike train distances}

In \citet{Satuvuori17} all three of the time-resolved distances described in Secs. \ref{App:ISI}-\ref{App:RI-SPIKE} have been adapted for data containing multiple time scales by adding a notion of the relative importance of local differences compared to the global time scales. The adaptive versions start to gradually ignore differences between spike trains for ISIs that are smaller than a minimum relevant time scale (MRTS). The MRTS is implemented by an additional parameter $\mathcal{T}$ which can either be set by the user or estimated directly from the data.

In the present study we basically evaluated these adaptive versions with the threshold parameter $\mathcal{T}$ set to zero, which is equivalent to using the original distances. This makes sense since the generalizations are primarily designed to reduce the importance of small time scales in datasets containing multiple time scales, but the steady rate Poissonians analyzed here contain only one time scale. 

The only reasonable comparison using the adaptive versions is when the threshold is fixed to a constant value. This causes higher rates to be considered as less significant for dissimilarity (once the spike trains get very dense, relative differences in ISIs or spike times hardly matter).
So for completeness, in Fig. \ref{Fig:A_ISI_SPIKE_3D} we provide the results for the adaptive distance to be compared with Fig. \ref{Fig:ISI_SPIKE_3D}.
Higher rates with a rate ratio close to 1 are considered as more similar because the differences are small in comparison to the threshold.
For a smaller threshold (Fig. \ref{Fig:A_ISI_SPIKE_3D}B) the area where similarity is enforced moves to higher frequencies.
The adaptive versions are designed to work with datasets containing multiple time scales such as regular spiking and bursts and for a fixed threshold the spike trains with a higher rate are considered as long bursts.

We also looked at the results for an automated threshold. Here the graphs look almost identical to the $\mathcal{T} = 0$ case (results not shown). However, and more importantly, the results can not really be compared in a meaningful way, since the threshold is different for different spike train pairs (see \citeauthor{Satuvuori17}, \citeyear{Satuvuori17}, for details). 

\section*{\label{s:References} References}

\end{document}